\documentstyle[11pt]{article}

\small
\begin{document}
\begin{titlepage}
\title{
{\bf Chiral phase transition in an expanding quark antiquark plasma } }
\author{ {\bf Abdellatif Abada}\thanks
{email : "ABADA@ptdec1.tphys.physik.uni-tuebingen.de"}
\thanks{New address:
Institute for Theoretical Physics, T\"ubingen University,
Auf der Morgenstelle 14, D-72076 T\"ubingen}
{ \bf and J\"org Aichelin}\thanks {email : "AICHELIN@nanvs2.in2p3.fr"}
\\
{\normalsize
Laboratoire de Physique Nucl\'eaire, CNRS/IN2P3 Universit\'e de Nantes, }\\
{\normalsize    2, Rue de la Houssini\`ere, 44072 Nantes Cedex 03, France }
}
\date{}
\maketitle
\begin{abstract}
We investigate the time evolution of a quark antiquark plasma by solving
numerically the relativistic transport equations
derived on the Hartree level from the Nambu-Jona-Lasinio model.
We find that the phase transition in the expanding quark antiquark plasma
is different as compared to that in a static plasma.
The expansion competes with the transition and finally quark droplets
will be formed which subsequently hadronizes.
These findings raise the question whether static thermal models
can make at all any prediction about signals of that transition .
\end{abstract}


\vskip 1cm
LPN 94-02 \hfill{December 1994}

\vskip 1cm
{\it  To be published in Phys. Rev. Lett.}
\end{titlepage}

\newpage
The search for the Quark Gluon Plasma (QGP)  is one of the
challenges of present day nuclear physics. It is the driving
motivation of the lead beam experiments which started this
year at CERN.
There is first of all the question whether this phase transition
can indeed be  achieved with accelerated nuclei. The even bigger
challenge is, however, to investigate which signals bear witness to the
existence
of the plasma, i.e., cannot be created by the dense hadronic medium
which is present after the QGP has ceased to exist and the quarks
are again bound in the hadrons which will finally be detected.

Up to now almost all investigations have tacitly assumed that the plasma
is contained in a box whose boundaries can be moved that slowly that
the systems is thermally equilibrated. Then,
knowing the equation of state, the thermodynamical variables
are sufficient to describe the transition to the hadronic world
(for a review of the thermal scenario we refer to \cite{PP}).
The reality, however, is most probably quite opposite. We have a gas of almost
massless
quarks at a temperature of more then 100 MeV and nothing which confines
the plasma to a certain space region.
Hence the system expands and we are confronted with the not yet investigated
question how the phase transition takes place in the expanding
non thermal system.

This question cannot be addressed in the framework of thermodynamics but
requires a transport theory. In principle a transport theory
may be derived from each Lagrangian and consequently also from
the QCD Lagrangian. The attempts which have been made, however, do not
come even close to an equation which can be used for numerical investigations
\cite{gy}. Therefore one has to rely on phenomenological Lagrangians. One of
these
is the Nambu-Jona-Lasinio Lagrangian (NJL)\cite{NJL}. In this Lagrangian
all gluonic degrees of freedom are integrated out.
The interaction between the quarks, however, is quite reliable
modeled, giving the
masses of the known hadrons with an acceptable precision. Its
drawback is that it has no confinement, i.e., cannot describe properly
the hadronization. Despite of this fact,
this Lagrangian seems to be appropriate
to start with the investigation of the question how a phase
transition takes place in an expanding plasma. The results can
be compared with that of the string models which recently also
conjectured that mini plasmas have to be formed during the expansion
of the system \cite{WA}. There is no doubt that this approach
can be regarded only as
a first step towards the understanding of the expansion of a quark
gluon plasma.

By using the loop expansion approach of the two-point source
connected generating functional \cite{CJT} of the NJL Lagrangian,
Zhang and Wilets \cite{ZW} derived a microscopic
transport theory to describe the chiral dynamics in high energy heavy-ion
collisions. This transport theory which describes the time evolution
of the Green's function is based on the
the closed time-path Green's functions formalism \cite{Ch}.
At this level it seems, however, not yet possible to perform calculations.
If one assumes that
the dissipation parts of the Green's functions are small compared to the
dispersive parts (the so-called quasiparticle limit), the quasiparticle energy
spectra of quarks and mesons are obtained from the equations giving the poles
of
the Green's functions, i.e., the zeros of the dispersive parts. Consequently,
the Green's functions representation can be converted into phase space
densities \cite{ZW}.

In this work we focus on the transport equation without collision terms.
For the NJL model, the Vlasov equation for the quark
density distribution $f({\bf r},{\bf p},t)$ in the Hartree level reads
as follow \cite{ZW,AA}
\begin{equation} \label{1}
{\displaystyle \frac{\partial f}{\partial t} +
\frac{{\bf p}}{E} . \mbox{\boldmath $\nabla_r$}f -
\mbox{\boldmath $\nabla_r$} E . \mbox{\boldmath $\nabla_p$} f = 0 }
\end{equation}
where $E$ is the relativistic quark energy $\sqrt{ {\bf p}^2 + M_c^2}$ and
$M_c$ is the constituent quark mass which is determined by the gap equation
\cite{ZW} :
\begin{equation}\label{2}
{\displaystyle
M_c({\bf r},t) = m + g^2M_c({\bf r},t)
\int_0^\Lambda \frac{\hbox{d}^3p}{ \sqrt{ {\bf p}^2 + M_c^2({\bf r},t)} }
(\frac{2 N_c N_f}{(2\pi)^3}  -
f({\bf r},{\bf p},t) - \tilde f({\bf r},{\bf p},t)~) }~.
\end{equation}
In Eq.(\ref{2}) $m$ is the quark mass ($m=m_u=m_d=4$ MeV), $N_c=3$ is the color
number, $N_f=2$ is the flavor number, $g$ is the
NJL four-point interaction coupling constant \cite{NJL} and
$\Lambda$ is the cutoff in momenta. In order to reproduce the pion mass and
the pion decay constant the parameters $g$ and $\Lambda$ are fixed
to 0.48 fm and 820 MeV respectively \cite{ZW}.

The solution of Eq.(\ref{1}) is not trivial due to the dependence of the
energy $E$ on the distribution $f$ via the constituent quark mass (c.f.,
Eq.(\ref{2})). For this reason we have to use some approximation scheme.
In this work we
use the test particles method which has been shown to be powerful in
intermediate energy heavy ion collisions \cite{WBKD}.
This approach consists to approximate
the distribution function $f$ ($\tilde f$) by a crowd of "numerical" quarks
(anti-quarks). Namely,
\begin{equation} \label{3} \begin{array} {ll}
{\displaystyle f ({\bf r},{\bf p},t) = w \sum_{i=1}^{A}
\delta^3 ({\bf r}-{\bf r}_i(t)) ~\delta^3({\bf p}-{\bf p}_i(t)) } \\
{\displaystyle \tilde f ({\bf r},{\bf p},t) = w \sum_{i=1}^{\tilde A}
\delta^3 ({\bf r}-\tilde {\bf r}_i(t)) ~\delta^3({\bf p}-\tilde {\bf p}_i(t)) }
\end{array}  \end{equation}
where $A$ and $\tilde A$ are the number of test-particles and
anti-particles respectively, while
$w$ is a normalization factor which is related to the number $B$ of the real
quarks of system by
\begin{equation} \label{bar}
{\displaystyle
\int d^3r d^3p ~(f ({\bf r},{\bf p},t) -\tilde f ({\bf r},{\bf p},t) )
= w(A-\tilde A) \equiv B }~.
\end{equation}
To satisfy Eq.(\ref{1}), the position and momentum of the $i$-th numerical
quark (anti-quark) should obey the relativistic Hamilton's equations. Namely,
\begin{equation} \label{4} \begin{array} {ll}
{\displaystyle \dot {\bf r}_i = {\bf p}_i/E_i~~~,~~~
\dot {\bf p}_i = -\vec \nabla_{r} E_i~~~,~~~
E_i = \sqrt{ {\bf p}_i^2(t) + M_c^2({\bf r}_i,t)} }\\
{\displaystyle \dot {\tilde {\bf r}_i} = \tilde {\bf p}_i/\tilde E_i~~~,~~~
\dot {\tilde {\bf p}_i} = -\vec \nabla_{r} \tilde E_i~~~,~~~
\tilde E_i = \sqrt{ \tilde {\bf p}_i^2(t) + M_c^2(\tilde {\bf r}_i,t)} }
\end{array} \end{equation}
where dots indicate time differentiations.
We have investigated numerically the equations of motion (\ref{4}) together
with the gap equation (\ref{2}) in a self-consistent manner.
We have performed our calculations on
a 3 dimensions 41 point lattice with a $a =0.25$ fm mesh size.
The initial conditions are taken as follows : we have considered $A$
test-quarks and $\tilde A$ test anti-quarks
(see below for the numerical values) distributed randomly within a sphere
of radius $r_0=1.05$ fm. The position of each test
particle is assigned by choosing:
\begin{equation}
{\displaystyle r=r_0(x_1)^{1/3} ~~~;~~~\cos(\theta) = 1-2x_2~~~;~~~
\phi = 2\pi x_3 }
\end{equation}
where $x_1,x_2$ and $x_3$ are random numbers between 0 and 1. The Cartesian
coordinates of the test particle are $r\cos\phi \sin \theta$,
$r\sin\phi \sin \theta$ and $r\cos\theta $.
The momentum of the test particles are choosen randomly
according to a Fermi distribution. It means that initially the system is
in equilibrium. The coordinates of each test-particle momentum are
$[p_x = p\cos\phi_p \sin \theta_p,~ p_y= p\sin\phi_p \sin \theta_p,~
p_z = p\cos \theta_p]$
where the angles are choosen randomly as above while $p$ is determined from
\begin{equation}
{\displaystyle
\int_0^p p^{'2}dp' ~(1+e^{\frac{E_{p'}- \mu}{T}})^{-1} ~/
\int_0^{\Lambda} p^{'2}dp' ~(1+e^{\frac{E_{p'}- \mu}{T}})^{-1}
= x_1 }
\end{equation}
In the last equation $x_1$ is a random number,
$E_{p'}= \sqrt{ p^{'2} + M_c^2(0)}$ where $M_c(0)$ is
the space-independent solution of the gap equation (\ref{2}) for a fermi
distribution, $T$ represents the temperature, $\mu$ the chemical potential and
$\Lambda$ the cutoff. The minus sign in the exponentials should be replaced
by a plus sign in the case of a test anti-particle.
In this work we have taken initial quark and energy densities
of $\rho_0 = B/\frac{4}{3} \pi r_0^3 = 1.87$ fm$^{-3}$ (which
leads to $B=9$ for $r_0=1.05$ fm ), and
$e_0$ = 1.57~Gev~fm$^{-3}$ respectively. These values correspond, via the
definition of quark and energy densities  in the presence of a cutoff
\begin{equation}\begin{array} {ll} \label{rhoe0}
\rho_0 = &{\displaystyle
\frac{2N_cN_f}{(2\pi^3)} \int^{\Lambda} d^3p
\left( \frac{1}{ 1+e^{(E_p-\mu)/T}} - \frac{1}{ 1+e^{(E_p+\mu)/T}} \right) }\\
e_0 = &{\displaystyle
\frac{2N_cN_f}{(2\pi^3)} \int^{\Lambda} d^3p ~E_p
\left( \frac{1}{ 1+e^{(E_p-\mu)/T}} + \frac{1}{ 1+e^{(E_p+\mu)/T}} \right) }
\end{array}\end{equation}
together with the gap equation (\ref{2}), to the following values for the
temperature, the chemical potential and the constituent quark mass
respectively:
$ T=240$ ~MeV, $\mu = 200$ ~MeV and  $M_c(0) = 33.5$ ~MeV.
The numerical values of $A$ and $\tilde A$ for the
test-quarks and test anti-quarks respectively have been fixed by fitting,
at $t=0$, to  $\rho_0$ and $e_0$ given above with the result:
$A=2630$ and $\tilde A=630$. According to Eq.~(\ref{bar})  the difference
$A-\tilde A$ determines how many test (anti) quarks one employs for
a physical (anti) quark \cite{foot}.
At this stage we want to stress that the well known expressions of the quark
 and  energy densities in the case of a fermi distribution
with a vanishingly small quark mass:
\begin{equation}\begin{array}{ll}
\rho_0 = N_c N_f \mu (\pi^2 + (\beta\mu)^2)/3\pi^2\beta^2
\\
e_0 = N_c N_f  (21 \zeta(4) + 6 (\beta\mu)^2 \zeta(2) + \frac{1}{2}
(\beta \mu)^4)/2\pi^2\beta^4  ~,
\end{array}\end{equation}
lead to $T=1/\beta=185$ MeV and $\mu=190$ MeV. The difference between these
values and those given above is due to the cutoff.
To be consistent we have  considered in this work the
quantities calculated with the cutoff $\Lambda$ (c.f., Eqs. (\ref{rhoe0}) ).

For the numerical implementation of our equations, the evolution of the
$i$-th test particle or test anti-particle (c.f., Eqs(\ref{4})) is represented
by
the following algorithm
\begin{equation}\label{6} \begin{array}{ll}
{\displaystyle
{\bf p}_i(t+\delta t) = {\bf p}_i(t-\delta t) - 2 ~\delta t
(~\vec \nabla M_c({\bf r}_i,t) ) ~M_c({\bf r}_i,t)/E_i(t) }\\
{\displaystyle
{\bf r}_i(t+\delta t) = {\bf r}_i(t-\delta t) + 2 ~\delta t ~
{\bf p}_i(t)/E_i(t) }
\end{array}\end{equation}
where $\delta t=0.05$ fm/c is the mesh in time.
At the time $t$, we determine the
constituent mass $M_c(cell,t)$ (\ref{2}) at each cell of size $a$ of the
space lattice ($a$ being the space mesh-size) according to
\begin{equation} \label{7}
{\displaystyle
M_c(cell,t) = m + g^2 M_c(cell,t) ~(\frac{3}{2\pi^3} I_{\Lambda}(t) -
\frac{w}{a^3} \sum_{i=1}^{A_{cell}} \frac{1}{E_i(t)} -
\frac{w}{a^3} \sum_{ i =1}^{\tilde A_{cell}}
\frac{1}{\tilde E_i(t)} ) } \end{equation}
where $A_{cell}$ ($\tilde A_{cell}$) are all the test particles
(anti-particles) which are inside the considered cell, and
$I_{\Lambda}=\pi M_c^2(\sinh(2\alpha) - 2\alpha) $ with $\alpha =
\log [\frac{\Lambda}{M_c} + \sqrt{\frac{\Lambda^2}{M_c^2} +1} ~~]$.
In order to test the accuracy of our numerical method, we
have checked the energy conservation. We have found that the total energy of
the system changes by less than 5 percent for 40 iterations in time with a
0.05 fm/c mesh-time.

Before we come to the numerical results it is instructive to discuss our
expectations using a simplified model. Let us assume for the moment that
we have a freely expanding nonrelativistic plasma at a temperature T
which is initially confined in a sphere in coordinate space. The (Maxwell)
velocity distribution $dN/dv$ is zero for $v = 0$ and has a pronounced peak
around $ \bar v = \sqrt{2T/m}$. Without any interaction the many particles
with about that velocity stay together, i.e. the maximum of the density
move with $\bar v$ outwards. For radii considerably larger or smaller
than $r(t) =\bar v t, (t\gg 0)$ the density is lower. Thus the particles
are concentrated in a shell. The density in that shell decreases proportional
to $r^{-2}(t)$.

Now, what happens if we switch on the interaction. The interaction acts
like a repulsive momentum dependent potential $V=\alpha(\rho)\cdot g(p)$ where
$\alpha$ increases with decreasing density. Since the potential is more
repulsive in the low density region as compared to the high density
region there is a force which accelerates the particles towards
high density. Thus the high density zone can maintain a higher density
as compared to the free streaming on expense of the surrounding
lower density regions. Finally it is energetically favourable to
give up the isotropy in $\phi$ and $\theta$ and to build local fluctuations
(droplets) of high density which are separated by regions of lower density.
However, this is
beyond the limits of the predictive power of
our one body theory even if we observe these
fluctuations due to the finite number of test particles.

We come now to the numerical results. Fig.1 shows the behaviour
of the constituent quark mass $M_c$ as a function of the radius for different
times. For high density, i.e. in the plasma phase,
$M_c$ is equal to the bare mass of the quark. At zero density, after
the phase transition has taken place, $M_c$ is equal to the constituent quark
mass of $\approx 300$ MeV. Initially we see that all particles have
a mass of 33.5 MeV which can be considered as the bare mass.
The fact that it is not exactly equal to 4~MeV is
due to the finite density. Hence the system is in the quark phase.
Then the system expands and the mass of the particles close
to the surface comes close to the constituent quark mass.
At $t= 1$fm/c we see that
the particles have left the center, and therefore the mass has increased
also there to the constituent quark level. We have now an expanding
{\it shell}.
Before and after the shell we have constituent quarks while in the interior we
have
the plasma. During the expansion the peak density decreases
and therefore $M_c$ increases. The plasma in the interior of the shell
starts to make the phase transition. This phase transition is almost
accomplished after 1.9 fm/c.

Hence the result of the numerical calculations follows qualitatively
the consideration explained above. The following scenario of the
phase transition emerges. If initially a plasma is created in a
heavy ion collisions, it is, due to geometry, confined in a limited
space region. The high temperature of the plasma leads to a fast
expansion of the system. The phase transition takes place
while the system expands. There is a density front traveling outwards.
The density is fed by
the interaction of the quarks which pulls the quarks into the high
density region but it also decreases with about $r^{-2}(t)$ due to the
expansion. Most probably then the system gives up the spherical isotropy
and forms quark droplets. However, they will not be stable but will hadronize.
Obviously this scenario is quite different from that expected for
a system which keeps global thermal equilibration during the expansion.

Is this phase transition specific for the NJL Lagrangian or genuine for
a phase transition in an expanding system? A final answer is not at hand yet
but there exists another phase transition in nuclear physics which has
been investigated in detail recently \cite{gr} : The liquid (fragments)-gas
(nucleons)
transition during the expansion of a fireball created at much lower
energy ( $E_{\hbox{beam}} \sim$ 100 MeV/N) in heavy ion reactions. There,
for a density smaller than the normal nuclear matter density, we have
as well the situation that the potential is more attractive at higher
density as compared to lower density. In the expansion of this system a
similar droplet formation has been found in Ref. \cite{gr} if one describes
the expansion by a 1 - body transport theory as done here. Thus,
there is evidence that our results are rather general. More realistic
$N$-body calculations confirm this general structure and show that
indeed many droplets are formed.

Calculations including collision terms are under way. Recent
calculations of the hadronization cross section of a quark plasma \cite{hu}
offers as well the possibility to model the chiral phase transition itself.
The results will be reported in a forthcoming publication \cite{AA}.

\vskip .5cm
\noindent {\Large {\bf Acknowledgement} }
\vskip .5cm

We would like to thank Prof. H. Reinhardt for stimulating discussions about
the Nambu-Jona-Lasinio model.

\vskip .5cm
\noindent {\Large {\bf Figure captions } }
\vskip .5cm

{\bf FIG. 1.} The constituent quark mass $M_c$ (c.f.,
Eqs. (\ref{2} ,\ref{7}) ) as a function of the radius for different times.
The initial temperature and density of the system are $T = 240 $ MeV and
$\rho_0 = 1.87$ fm$^{-3}$ respectively.

\end{document}